\documentclass{article}
\usepackage[T1]{fontenc}
\usepackage[utf8]{inputenc}
\usepackage{ismir} 
\usepackage{amsmath,cite,url}
\usepackage{graphicx}
\usepackage{color}

\usepackage{microtype}

\usepackage{cite}
\usepackage{amsmath,amssymb,amsfonts}
\usepackage{graphicx}
\usepackage{xcolor}
\usepackage{textcomp}
\usepackage{subfigure}
\usepackage{enumitem}
\usepackage{multirow, array}
\usepackage{arydshln}
\usepackage{etoolbox,siunitx}
\usepackage{caption, subcaption, amsmath, graphicx, amssymb, changepage, cite}
\usepackage{algpseudocode}
\usepackage{balance}
\usepackage{booktabs}

\title{ITO-Master: Inference-Time Optimization for Audio Effects Modeling of Music Mastering Processors}

\multauthor
  {Junghyun Koo$^1$ \hspace{1cm} Marco A. Martínez-Ramírez$^1$ \hspace{1cm} Wei-Hsiang Liao$^1$}
  {{\bf Giorgio Fabbro$^2$ \hspace{1cm} Michele Mancusi$^2$ \hspace{1cm} Yuki Mitsufuji$^{1,3}$}\\
  $^1$ Sony AI, Japan \;
  $^2$ Sony Europe B.V., Germany \;
  $^3$ Sony Group Corporation, Japan \\ 
  {\tt\small \{firstname.lastname\}@sony.com}
  }

\def\authorname{J. Koo, M. Martínez-Ramírez, W-H. Liao, G. Fabbro, M. Mancusi, and Y. Mitsufuji}

\usepackage[bookmarks=false,pdfauthor={\authorname},pdfsubject={\pdfsubject},hidelinks]{hyperref}

\begin{document}

\maketitle

\begin{abstract}
Music mastering style transfer aims to model and apply the mastering characteristics of a reference track to a target track, simulating the professional mastering process. However, existing methods apply fixed processing based on a reference track, limiting users' ability to fine-tune the results to match their artistic intent. In this paper, we introduce the ITO-Master framework, a reference-based mastering style transfer system that integrates Inference-Time Optimization (ITO) to enable finer user control over the mastering process. By optimizing the reference embedding \( z_{\text{ref}} \) during inference, our approach allows users to refine the output dynamically, making micro-level adjustments to achieve more precise mastering results. We explore both black-box and white-box methods for modeling mastering processors and demonstrate that ITO improves mastering performance across different styles. Through objective evaluation, subjective listening tests, and qualitative analysis using text-based conditioning with CLAP embeddings, we validate that ITO enhances mastering style similarity while offering increased adaptability. Our framework provides an effective and user-controllable solution for mastering style transfer, allowing users to refine their results beyond the initial style transfer. 


\end{abstract}

\section{Introduction}
\label{sec:introduction}

Music mastering is the final step in the audio production process, ensuring professional sound quality and consistent playback across music distribution platforms. This process involves applying a series of audio effects such as equalization, compression, stereo imaging, and limiting, which collectively shape the sonic characteristics and enhance the overall quality of the audio~\cite{zolzer2002dafx, shelvock2012audio}. Traditionally, mastering has required skilled engineers who carefully adjust these effects based on the track’s content and desired artistic outcome. However, with the increasing volume of music production and the demand for consistency across streaming platforms, the need for automated mastering solutions has grown substantially.

In response to this demand, various automatic mastering systems have emerged~\cite{mastering_services,landr,ramirez2021differentiable}. However, these systems operate in an unconditioned manner, applying audio effects without direct user control. To introduce adaptability, reference-based approaches have been explored, where the processing characteristics of a reference track are applied to another~\cite{mee, matchering}. These methods aim to match audio features such as dynamics, tonal balance, and stereo width, offering an alternative to fully automatic mastering. However, significant challenges remain in achieving both high-quality results and controllability. 

Existing reference-based approaches can be broadly categorized into \textit{black-box} and \textit{white-box} models. Black-box models, often based on end-to-end neural networks~\cite{mee}, can effectively capture high-level audio patterns but lack transparency and interpretability, making it difficult for users to modify specific aspects of the processing. In contrast, white-box models leverage either feature matching algorithms~\cite{matchering} or differentiable audio processors~\cite{ramirez2021differentiable} to provide greater control over individual parameters. While white-box methods offer a structured and interpretable approach~\cite{ddsp}, they are often constrained by the simplicity of their differentiable processors, which may not fully replicate the complex tools used in professional mastering.

In this paper, we introduce the ITO-Master framework, a reference-based audio effects modeling of music mastering processors that incorporates Inference-Time Optimization (ITO) for finer user control. While previous style transfer methods apply fixed processing based on a reference, our approach allows users to dynamically refine the output when the initial result does not fully align with their preferences. By optimizing the reference embedding \( z_{\text{ref}} \) during inference, ITO-Master enables micro-level adjustments, allowing for more precise and targeted mastering refinements.

The key contributions of this work include: 
(1) \textbf{ITO-Master framework}: A novel approach for reference-based audio effects modeling of mastering processors using ITO.
(2) \textbf{Comparison of black-box and white-box models}: A systematic study of two paradigms for evaluating their effectiveness and trade-offs.
(3) \textbf{Realistic mastering processor chain}: Implementation of a structured differentiable mastering pipeline to enhance the realism of white-box processing.
(4) \textbf{Comprehensive evaluation}: Performance validation via objective metrics, listening tests, and qualitative analysis using text-based conditioning with CLAP embeddings.

\begin{figure*}[t]
    \centering
    \subfigure[Training pipeline of Mastering Style Converter \( \Psi \). During this phase, Reference Encoder \( \Phi \) is trained using diverse mastering styles generated by random FX manipulation \( f \). The target signal \( y \) is synthesized by applying the same manipulation to segment \( A \) as to reference segment \( B \), both from the same song.]{
        \includegraphics[width=0.55\linewidth]{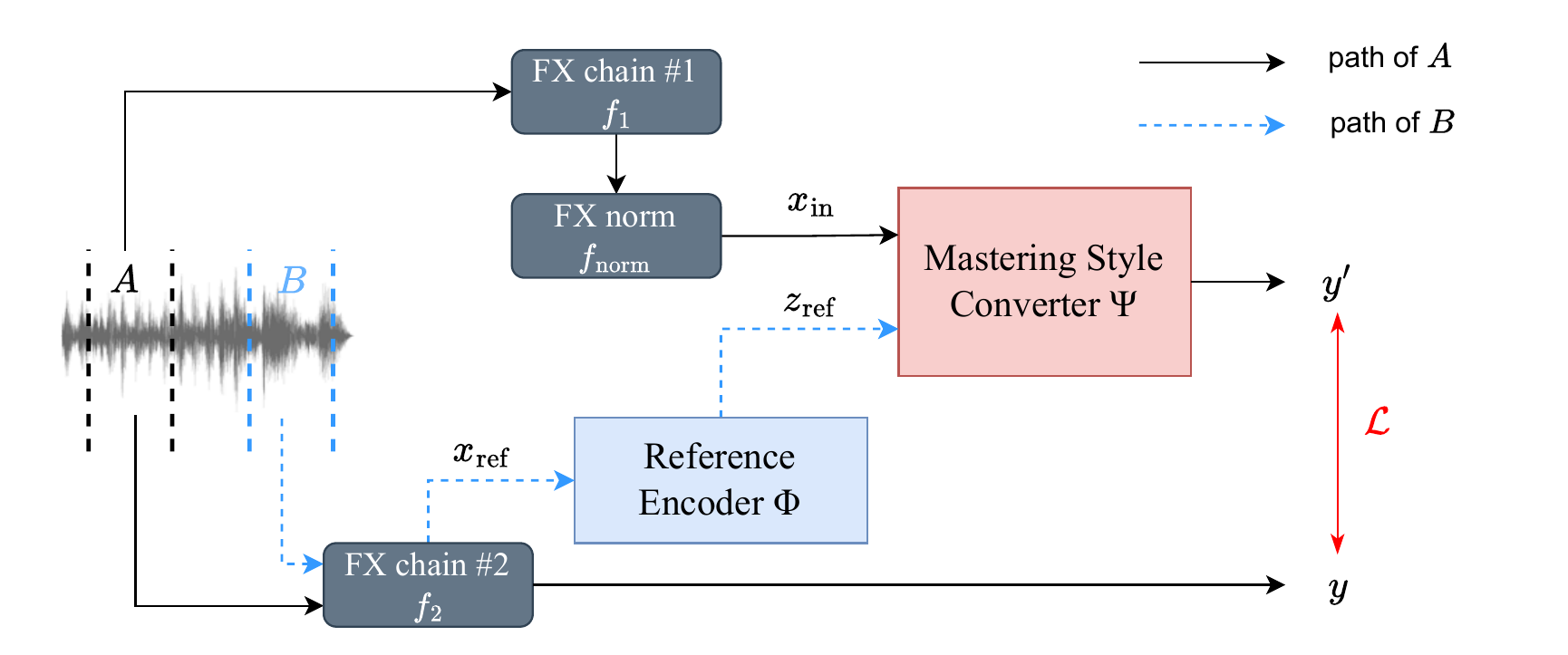}
        \label{subfig:training_converter}
    }
    \hspace{0.5cm}
    \subfigure[ITO is performed using an auxiliary (content-independent) objective function, allowing any reference music. Users can optimize \( z_{\text{ref}} \) based on their preferences for both the reference and objective function.]{
        \hspace{0.3cm}
        \includegraphics[width=0.35\linewidth]{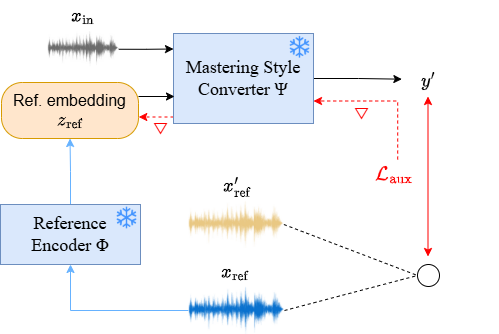}
        \label{subfig:training_ITO}
    }
    \vspace{-0.3cm}
    \caption{Overall pipeline of ITO-Master.}
    \vspace{-0.2cm}
    \label{fig:training_pipeline}
\end{figure*}

\section{Related Works}
\label{sec:related_works}

\subsection{Audio Effects Style Transfer}
\label{subsec:FX_style_transfer}
Audio effects style transfer has become a significant area of research in automating and enhancing music production. Recent advancements in deep learning have led to more sophisticated approaches, where neural networks are used to learn complex mappings between input and output audio signals. These methods have been applied to style transfer across single audio effects (Fx) or multiple sets of Fx (Fx chain)~\cite{ramirez2021differentiable, mee, reverb_conversion, lee2023blind, DeepAFx-ST, MST, diff-MST, chen2024towards}, effectively modeling temporal dependencies and applying style transfer based on a reference track at the waveform level. While these methods have shown success in controlled environments, challenges remain in extending their application to diverse real-world scenarios, particularly in adapting to different mastering styles and varying input signals.

\subsection{Inference-Time Optimization}
\label{subsec:ITO_related}
Recently, \cite{ditto, ditto2} have explored inference-time optimization (ITO) in music generation tasks, where the initial latent embedding is optimized by backpropagating through diffusion-based models with the loss between a generated sample and a reference track. In the context of audio effects style transfer, ITO has been applied in methods like ST-ITO~\cite{ST-ITO, yu2025improving}, where interpretable parameters in a white-box differentiable Fx chain are optimized.

Our work focuses specifically on mastering style transfer, where handling heavily compressed audio—common in commercially released music—requires particular attention to limiters and dynamic range management. The ITO-Master framework introduces ITO on the reference embedding \( z_{\text{ref}} \), allowing optimization at inference time in both black-box and white-box models. By fine-tuning \( z_{\text{ref}} \), our approach adapts the mastering style of a reference track without retraining the entire model. ITO-Master ensures smooth adaptation to the reference track's characteristics while preserving the ability to observe and adjust the underlying parameters in the white-box model. This makes our framework particularly suitable for mastering tasks, providing professionals and amateurs with precise control over the audio mastering process.

\section{Methodology}
\label{sec:method}
In this section, we describe the components of the proposed mastering style transfer framework: the training pipeline, Mastering Style Converter, differentiable mastering chain, and ITO. The training pipeline simulates real-world mastering scenarios by applying random Fx manipulations. The Mastering Style Converter \( \Psi \) transfers the mastering style from a reference track to a target track and can be implemented using both black-box and white-box approaches. The differentiable mastering chain serves as a white-box processor that models various Fx in a structured sequence. Lastly, the ITO process optimizes \( z_{\text{ref}} \) at inference time to enhance style transfer performance. The following subsections describe each component in detail.

\subsection{Training Pipeline of Mastering Style Transfer}
\label{subsec:training_pipeline}
As shown in Figure~\ref{subfig:training_converter}, the training pipeline follows established methodologies in style transfer, utilizing a self-supervised training framework with random Fx manipulation~\cite{mee,DeepAFx-ST,MST,diff-MST}. Based on the understanding that a single song maintains a consistent mastering style throughout~\cite{shelvock2012audio}, a song is first segmented into two parts, \( A \) and \( B \). For the input to \( \Psi \), random manipulation \( f_1 \) is applied to simulate a random style, similar to how the process would function in an application setting. Then, we apply Fx-Normalization~\cite{FXnorm} \( f_{\text{norm}} \), which normalizes certain Fx characteristics to fixed target levels, to facilitate the performance of style transfer. \( f_{\text{norm}} \) is only applied to the equalizer (EQ), stereo imager, and loudness levels, allowing the model to capture a broader range of nonlinear FX transformations, such as compression levels and distortion. While normalizing compression is technically feasible, it is not well-suited for the on-the-fly training procedure. For distortion, normalization would require either removing all distortion from the given song or applying a consistent distortion level across all tracks, which is outside the scope of this paper. In summary, the input to \( \Psi \) is defined as \( x_{\text{in}} = f_1(f_{\text{norm}}(A)) \).

To achieve style transfer, the Fx information from the reference track \( x_{\text{ref}} \) is encoded to create the reference embedding that conditions \( \Psi \). A second random manipulation \( f_2 \) is applied to segment \( B \), which is then encoded by the reference encoder \( \Phi \), resulting in the reference embedding \( z_{\text{ref}} = \Phi(f_2(B)) \). The training process minimizes the loss between the model output \( y' = \Psi(x_{\text{in}}, z_{\text{ref}}) \) and the target signal \( y = f_2(A) \). Since both \( A \) and \( B \) originate from the same mastered song, we assume that \( y \) and \( x_{\text{ref}} \) share the same mastering style.

\subsection{Mastering Style Converter}
\label{subsec:converter}
\( \Psi \) can be implemented using two distinct modeling approaches: black-box modeling and white-box modeling. In the black-box approach, $\Psi$ directly models the waveform signal \( y' \). Conversely, the white-box approach estimates the parameters \( \Theta \) for the differentiable mastering Fx chain. The formulation of the differentiable mastering Fx chain in the white-box model is identical to that used in the Fx manipulator $f$ for processing randomly mastered audio.

The training objective for \( \Psi \) is the multi-scale spectral loss \( \mathcal{L}_{\text{MSS}} \), applied to both left-right and mid-side channels (where mid = left + right, side = left - right), as used in~\cite{MST}.

\subsection{Differentiable Mastering Chain Modeling}
\label{subsec:fx_manipulation}
The mastering chain is designed to be fully differentiable white-box processor, serving a dual purpose: functioning both as a mastering style converter and as a random mastering manipulation module for training the converter. By modeling a wide range of variability in mastering styles, the chain enables the system to robustly handle and replicate the complexities of real-world music mastering. To simulate a realistic music mastering process~\cite{mastering_services}, the chain includes six distinct Fx modules: 1. \textit{6-band parametric equalizer}, 2. \textit{distortion}, 3. \textit{3-band compressor}, 4. \textit{makeup gain}, 5. \textit{stereo imager}, and 6. \textit{limiter}. 
The order of these modules is fixed, with the probability of applying each Fx module for random manipulation during training set at 90\%, 30\%, 80\%, 85\%, 60\%, and 100\%, respectively. These probabilities are adopted to introduce greater variability while preventing the synthesis of overly unrealistic mastering styles to enhance \( \Psi \)'s modeling capability. The chain comprises a total of 46 controllable parameters.

To ensure differentiability, the Fx modules are implemented using open-source libraries\footnote{\url{https://github.com/csteinmetz1/dasp-pytorch}}$^{,}$\footnote{\url{https://github.com/DiffAPF/torchcomp}} that support gradient-based optimization techniques. For modeling the 3-band multiband compression, a fourth-order Linkwitz-Riley crossover filter~\cite{crossover} is first applied to split the signal into three bands, followed by a differentiable all-pole filter. A key component in the chain is the use of these differentiable all-pole filters, as modeled by~\cite{yu2024differentiable}, which enable the computation of both compression and expansion effects in both the multiband compressor and limiter. This capability is crucial for practical applications, allowing the system to manage both limiters and delimiters, which is particularly important in real-world scenarios where most commercially released music is heavily compressed~\cite{shelvock2012audio}, requiring effective style transfer under such conditions.

\subsection{Inference-Time Optimization on Reference Embedding}
\label{subsec:ITO_method}
The primary contribution of this work is the introduction of ITO on \( z_{\text{ref}} \). Instead of fine-tuning the entire model \( \Psi \), the focus is on optimizing only \( z_{\text{ref}} \) while keeping the pre-trained \( \Psi \) fixed, as shown in Figure~\ref{subfig:training_ITO}. Although optimizing \( z_{\text{ref}} \) during inference time in the Black-box model does not provide interpretability in terms of Fx processors, the White-box model preserves interpretability. In fact, the changes in the parameters \( \Theta \) before and after the ITO process can be observed, providing insight into how \( \Theta \) is adjusted. Additionally, users can optimize the system with an alternative reference signal \( x'_{\text{ref}} \), offering a different approach from conventional mastering style transfer, as this method merges mastering styles based on the combination of the new reference signal and the optimizing objective function. The advantages of ITO on \( z_{\text{ref}} \) include a significantly reduced number of optimization steps compared to optimizing the entire differentiable chain's \( \Theta \) from scratch, as we compare in Section~\ref{sec:results}.

For ITO, the Audio Feature (AF) loss proposed by~\cite{diff-MST} is utilized as the auxiliary objective function \( \mathcal{L}_\text{aux} \). The AF loss is a content-independent loss that combines various audio feature transformations, capturing the dynamics, spatialization, and spectral characteristics of the audio. Each transformation in the AF loss has its own predefined weighting factor, and these weighted transformations are summed together to compute the overall loss. In our experiments, the original weights from the reference paper are followed. We optimize \( z_{\text{ref}} \) iteratively using gradient descent: \( z_{\text{ref}}^{(t+1)} = z_{\text{ref}}^{(t)} - \eta \nabla_{z} \mathcal{L}_\text{aux}(\Psi(x_\text{in}, z_{\text{ref}}^{(t)}), x_\text{ref}) \), where \( \eta \) is the learning rate. For objective and subjective evaluation, we focus solely on using AF loss as the optimization objective for ITO. 
Since ITO can be optimized with any loss function, we also qualitatively explore optimization using a text prompt with CLAP embeddings~\cite{Laion-CLAP} in Section~\ref{subsec:qual_analysis}.

\section{Experiments}
\label{sec:experiments}

\subsection{Dataset}
\label{subsec:dataset}
We utilized MoisesDB dataset~\cite{moisesdb} for training, and validated using the MUSDB18 validation subset~\cite{musdb18-hq}. 
Mixture samples from these datasets are employed, as they are not fully mastered, allowing random Fx manipulation with \( f \) to create synthetic mastered samples.
For Fx-Normalization, mean statistics are precomputed on the MoisesDB dataset, and normalization is applied to match the EQ, stereo imager, and loudness levels.
For evaluation, 200 songs are randomly selected from the MTG-Jamendo dataset~\cite{mtg-jamendo}. Of these, 100 songs are used as \( x_{\text{in}} \), and the remaining 100 serve as \( x_{\text{ref}} \) for input into \( \Psi \).
During training and validation, both segments \( A \) and \( B \) are 11.8 seconds long. For evaluation, 30-second samples are used since the fully convolutional architecture of \( \Psi \) can handle variable-length inputs.

\subsection{Experimental Setup}
\label{subsec:exp_setup}
The experimental setup includes two primary training configurations for \( \Psi \): Black-box and White-box methods. Both configurations use the Temporal Convolutional Network (TCN)~\cite{TCN} architecture for processing \( x_{\text{in}} \), with 10.5 million trainable parameters. Pre-trained weights of the \textit{FXencoder}~\cite{MST} is adopted as $\Phi$ and tested under two conditions: with and without \( \Phi \) being trained alongside \( \Psi \). All models are trained for 72,000 iterations with a batch size of 4.

In addition to training the \( \Phi \) and \( \Psi \) models, ITO is performed on each test data to optimize the reference embedding \( z_{\text{ref}} \), which has a dimensionality of 2048. The ITO process is run for a maximum of 100 steps or is stopped earlier if the loss value increases, indicating convergence. For comparison, an alternative ITO approach is also applied where optimization is performed solely on the parameters \( \Theta \) of the differentiable mastering chain \( f \). In this case, \( \Theta \) is optimized for up to 2K steps to evaluate its effectiveness relative to the proposed ITO method focused on \( z_{\text{ref}} \). All methods are optimized using the RAdam optimizer~\cite{radam} with a learning rate of $2\cdot 10^{-4}$.
More details are available in our open-source repository\footnote{\url{https://github.com/SonyResearch/ITO-Master}}.

\begin{table*}[t]
\centering
\renewcommand{\arraystretch}{1.3}
\resizebox{0.95\linewidth}{!}
{
\begin{tabular}{cl | >{\centering\arraybackslash}p{1.6cm}>{\centering\arraybackslash}p{1.6cm}>{\centering\arraybackslash}p{1.6cm}>{\centering\arraybackslash}p{1.6cm}>{\centering\arraybackslash}p{1.6cm}>{\centering\arraybackslash}p{1.6cm}}
\toprule
\multicolumn{2}{c|}{\textbf{Method}} & \textbf{AF ($\downarrow$)} & \textbf{DRV ($\downarrow$)} & \textbf{cos sim ($\uparrow$)} & \textbf{FAD$_\text{CLAP}$($\downarrow$)} & \textbf{FAD$_\text{DAC}$($\downarrow$)} & \textbf{FAD$_\text{EnCodec}$($\downarrow$)}  \\ \hline
\multirow{2}{*}{\textbf{Feature Matching}} 
                            & Fx-Normalization~\cite{FXnorm} & 0.157       & 0.801        & 0.941            & 161.4                 & 177.4                & 84.53 \\
                            & Matchering~\cite{matchering}       & 0.160       & 0.823        & 0.942            & 110.8                 & 126.1                & 59.34 \\ 
\midrule
\textbf{Baseline}           & E2E Remastering~\cite{mee}  & 0.288       & 0.858        & 0.942            & 104.3                 & 176.7                & 37.19 \\ 
\midrule
\multirow{7}{*}{\textbf{Proposed}}   
                            & Black-box        & 0.346       & 0.685        & 0.944            & 160.8                 & 378.4                & 51.12 \\
                            & \ \ \ + train $\Phi$    & 0.125       & 0.577        & 0.945            & 159.8                 & 177.4                & 46.94 \\
                            & \ \ \ \ \ \ + ITO on $z_\text{ref}$            & \textbf{0.099}       & 0.567        & \textbf{0.946}            & 182.2                 & 180.5                & 42.82 \\ \cdashline{2-8}
                            & White-box   & 0.253       & 0.598        & \textbf{0.946}            & 93.7                  & 144.8                & \textbf{36.22} \\
                            & \ \ \ + train $\Phi$    & 0.186       & 0.521        & 0.945            & \textbf{93.2}                  & \textbf{101.4}                & 38.90 \\
                            & \ \ \ \ \ \ + ITO on $z_\text{ref}$           & 0.139       & \textbf{0.474}        & \textbf{0.946}            & 105.2                 & 109.1                & 42.99 \\ \cdashline{2-8}
                            & ITO on $\Theta$         & 0.250       & 0.609        & 0.927            & 216.8                 & 294.8                & 101.60 \\ 
\bottomrule
\end{tabular}
}
\vspace{-0.2cm}
\caption{Mastering Style Transfer on Jamendo dataset (real-world scenario).}
\vspace{-0.3cm}
\label{tab:main_result}
\end{table*}

\subsection{Evaluation Metrics}
\label{subsec:eval_metrics}
To objectively evaluate mastering style transfer, content-independent objectives are utilized, given that different content is being compared. The following metrics are employed:

\begin{itemize}[leftmargin=*,itemsep=0pt,topsep=2pt]
    \item \textbf{Audio Feature (AF) Loss}: As discussed in Section~\ref{subsec:ITO_method}, AF Loss measures how well the output \( y' \) matches the desired audio features. 

    \item \textbf{Dynamic Range Variability (DRV)}: DRV is a crucial metric for assessing the compression level of audio, particularly in the context of music mastering, where limiters play a significant role. The DRV metric is computed by first identifying peak values in the audio signal using a high-frequency content onset detection function. DRV is the standard deviation of these peak values after filtering out the lowest 25\% of the values, which is defined as:
    
    \vspace{-0.7cm}
    \begin{equation}
    \text{DRV} = \frac{1}{C} \sum_{c=1}^{C} \text{std}\left(\left\{ p_i^c : p_i^c > \text{percentile}(p^c, 75) \right\}\right)
    \end{equation}
    
    where \( p^c \) denotes the set of peak values $\{ p_1^c,p_2^c,... \}$ in channel \( c \), and \( C \) is the total number of channels. The metric reflects the variability in dynamic range, with higher values indicating less consistent compression.

    \item \textbf{Fx Embedding Similarity (cos sim)}: Cosine similarity measures the similarity between the reference embedding \( \Phi(x_{\text{ref}}) \) and the output embedding \( \Phi(y') \). We adopt the pretrained FXencoder as \( \Phi \). This metric evaluates how closely the output matches the reference in terms of its learned Fx characteristics.

    \item \textbf{Fréchet Audio Distance (FAD)}: FAD~\cite{fad} assesses the perceptual quality of the generated audio by comparing the statistical distribution of the model outputs to a reference distribution. FAD is calculated using three deep audio embeddings: CLAP~\cite{MS-CLAP}, DAC~\cite{dac}, and EnCodec~\cite{encodec}. We chose these embeddings as they have shown strong correlation with human preference in acoustic quality, with codec-based models like DAC and EnCodec being particularly sensitive to acoustic effects, as demonstrated by~\cite{fad_analysis}. The metric computes the distance between the distribution of features extracted from the model’s output and those extracted from a subset of the Jamendo dataset, measuring how natural the style-transferred audio sounds compared to real recordings.
    
\end{itemize}

\subsection{Baseline Methods}
\label{subsec:baselines}
The following baseline methods, representing existing mastering style transfer systems, are used for comparison:

\begin{itemize}[leftmargin=*,itemsep=0pt,topsep=2pt]
    \item \textit{Acoustic Feature Matching Approaches}: \textbf{Feature matching} approaches aim to adjust the Fx of an input track to match those of a reference track by directly aligning specific audio features.
    \begin{itemize}[leftmargin=*,itemsep=0pt,topsep=2pt]
        \item \textbf{Fx-Normalization}~\cite{FXnorm}: Instead of normalizing the given audio to the mean statistics of the target data distribution, this approach directly matches the Fx levels to those of the reference song. The official implementation\footnote{\url{https://github.com/sony/FxNorm-automix}} is used to match the audio effects in the order of EQ, compression, stereo imaging, and loudness, respectively.
        \item \textbf{Matchering}~\cite{matchering}: An open-source library that matches the given song's RMS, frequency response, peak amplitude, and stereo width to those of the reference track. The official implementation\footnote{\url{https://github.com/sergree/matchering}} is used to infer the processed songs.
    \end{itemize}
    
    \item \textbf{E2E Remastering}~\cite{mee}: This end-to-end remastering approach is a black-box model that directly predicts the signal \( y' \) at the waveform level. The model is trained in a self-supervised manner using a large dataset of released pop songs. It leverages a pre-trained encoder and a projection discriminator to encourage the generation of realistic audio that accurately reflects the mastering style of the reference track.
\end{itemize}

\section{Results}
\label{sec:results}

\subsection{Objective Evaluation}
\label{subsec:obj_eval}
The performance of the proposed methods, along with the baseline approaches, is summarized in Table~\ref{tab:main_result}. 
The feature matching methods, specifically Fx-Normalization and Matchering, demonstrate strong performance on the AF and FAD metrics. This is expected, as these approaches directly apply Fx-related transformations to match the reference track's characteristics. However, these methods perform poorly on the DRV metric, as they lack the control needed for proper dynamic range adjustments, which is crucial in real-world mastering tasks involving delimiting.
E2E Remastering~\cite{mee} shows good performance on the FAD with CLAP and EnCodec embeddings, likely due to its use of an adversarial objective during training, which aids in generating realistic audio that closely matches the reference distribution. However, this system falls short on AF and DRV, indicating challenges in capturing precise audio feature transformations and managing dynamic range.

Among the proposed methods, Black-box approaches outperform White-box methods in terms of AF, indicating its effectiveness in capturing audio feature transformations from direct modeling of \( y' \) with \( \mathcal{L}_\text{MSS} \). However, the White-box method shows better results across all FAD metrics, suggesting that it produces audio more aligned with real-world distributions. This may imply that while Black-box models capture more detailed transformations, White-box approaches produce outputs that are more perceptually consistent with real-world mastering styles.

When \( \Psi \) is trained while keeping the pre-trained FXencoder \( \Phi \) fixed, the performance is generally inferior. This is likely because the FXencoder was trained on a different set of Fx chains and may not fully capture the manipulations applied to the reference songs in this context. However, when \( \Phi \) is trained alongside \( \Psi \), there is a significant improvement in performance, as this joint training allows the encoder to better adapt to the specific Fx manipulations used, leading to more accurate mastering style transfer.

Applying ITO on \( z_\text{ref} \) enhances AF performance, but introduces a trade-off in FAD scores, indicating while ITO can refine mastering style transfer, the number of optimization steps must be carefully calibrated to balance competing objectives. Conversely, applying ITO directly on \( \Theta \) yields poor results across all metrics, even with a larger number of optimization steps. Interestingly, in reverse engineering tasks---where the input and output content are identical---optimizing \( \Theta \) works well despite the complexity of the Fx chain~\cite{grafx, lee2025reverse, diffvox}. However, in mastering style transfer, content-independent loss functions are used to replicate only the mastering style from the reference track. This distinction highlights why ITO on \( \Theta \) is less effective in this context. Since \( \Psi \) is trained with a content-dependent objective, it leverages content information to enhance mastering style transfer. In contrast, $\mathcal{L}_\text{aux}$ used in ITO fails to capture the intricacies of the task, making it unsuitable for optimizing the entire differentiable chain in this scenario.

Audio samples are available on our demo page\footnote{\url{https://tinyurl.com/ITO-Master}}.

\begin{figure}[t]
    \centering
    \includegraphics[width=0.48\textwidth]{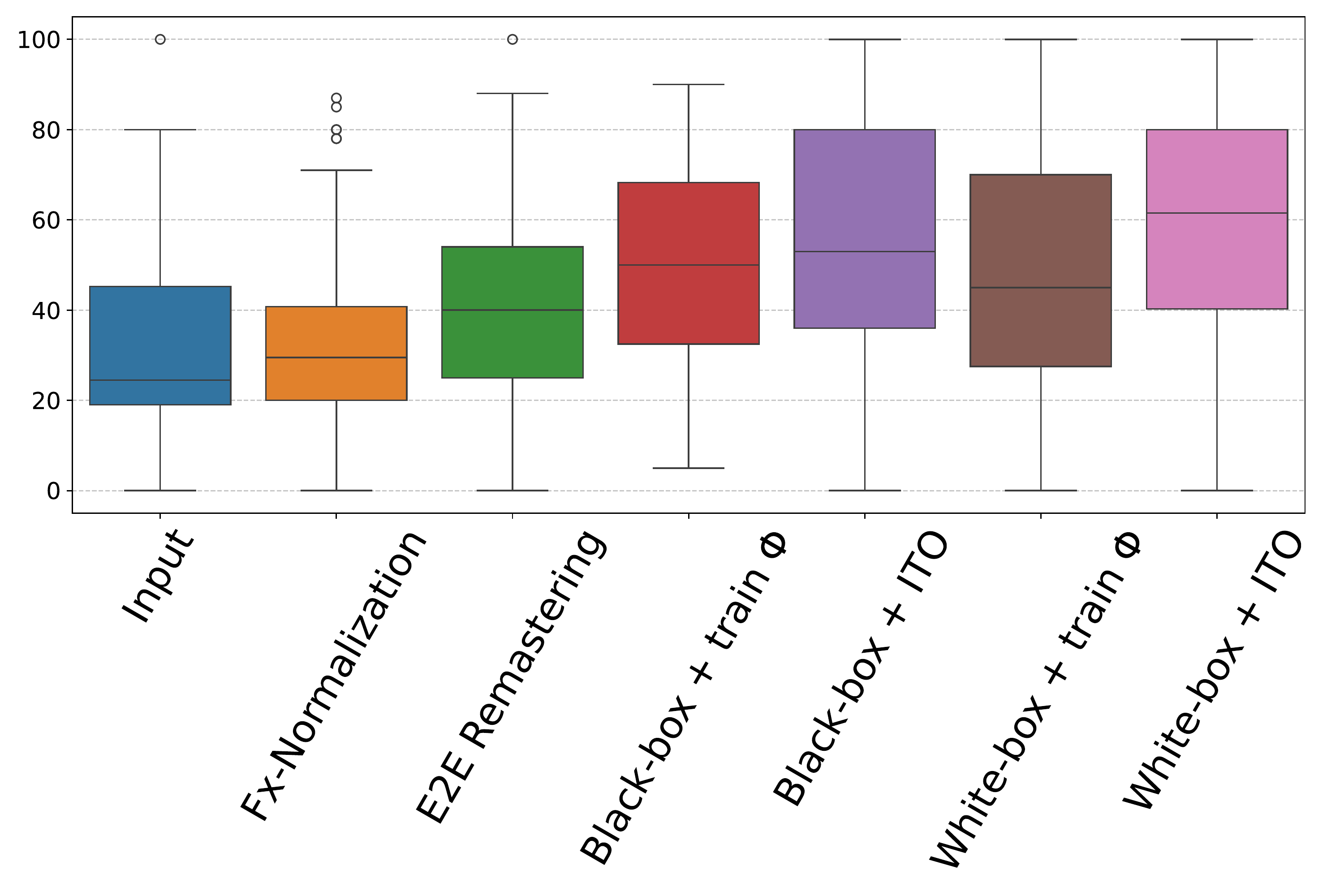}
    \vspace{-0.8cm}
    \caption{Subjective evaluation results.}
    \vspace{-0.4cm}
    \label{fig:subjective_result}
\end{figure}

\begin{figure*}[t]
    \centering
    \subfigure[Input Spectrogram]{
        \includegraphics[width=0.28\textwidth]{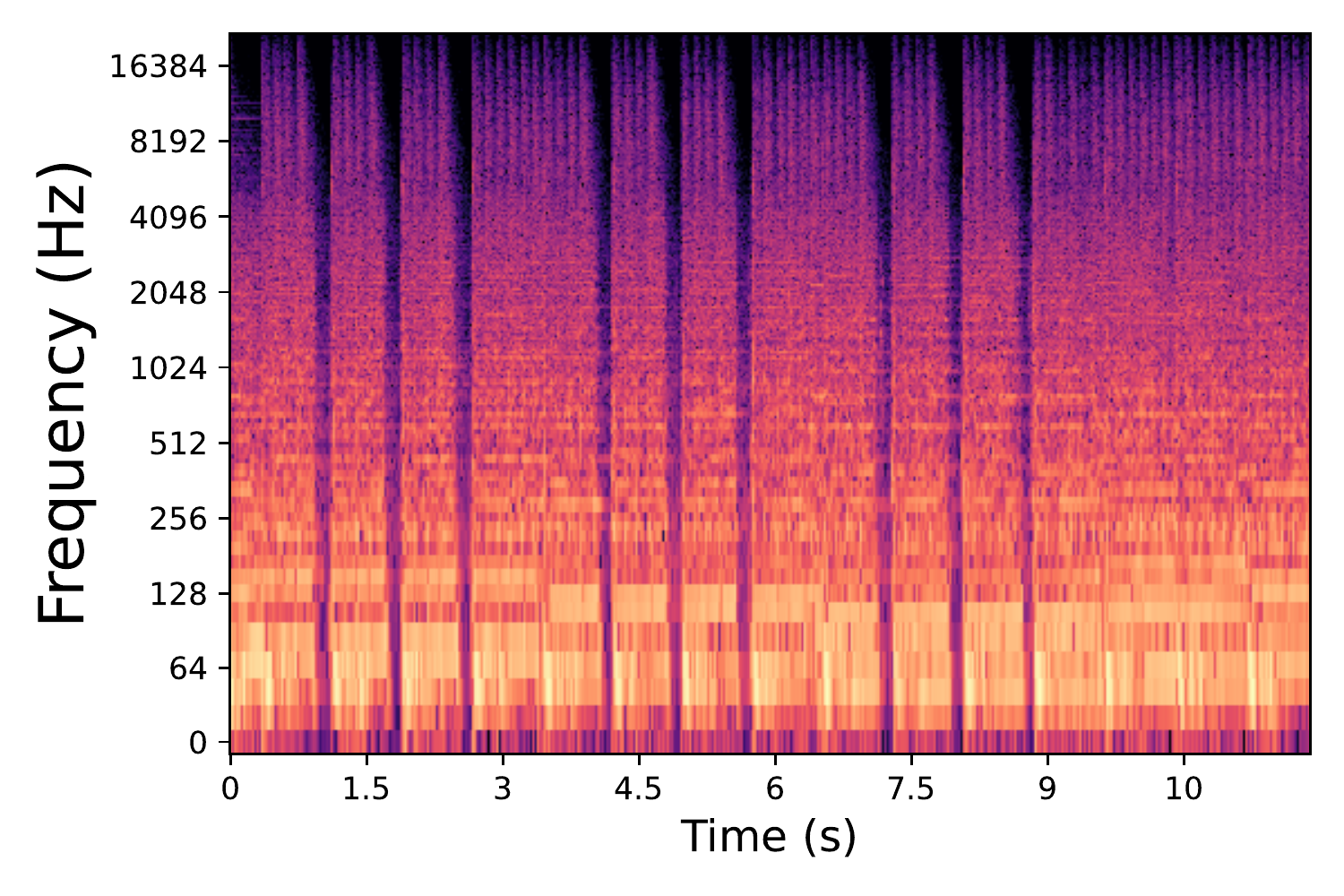}
    }
    \hfill
    \subfigure[Spectrogram Difference]{
        \label{subfig:qual_spec_diff}
        \includegraphics[width=0.66\textwidth]{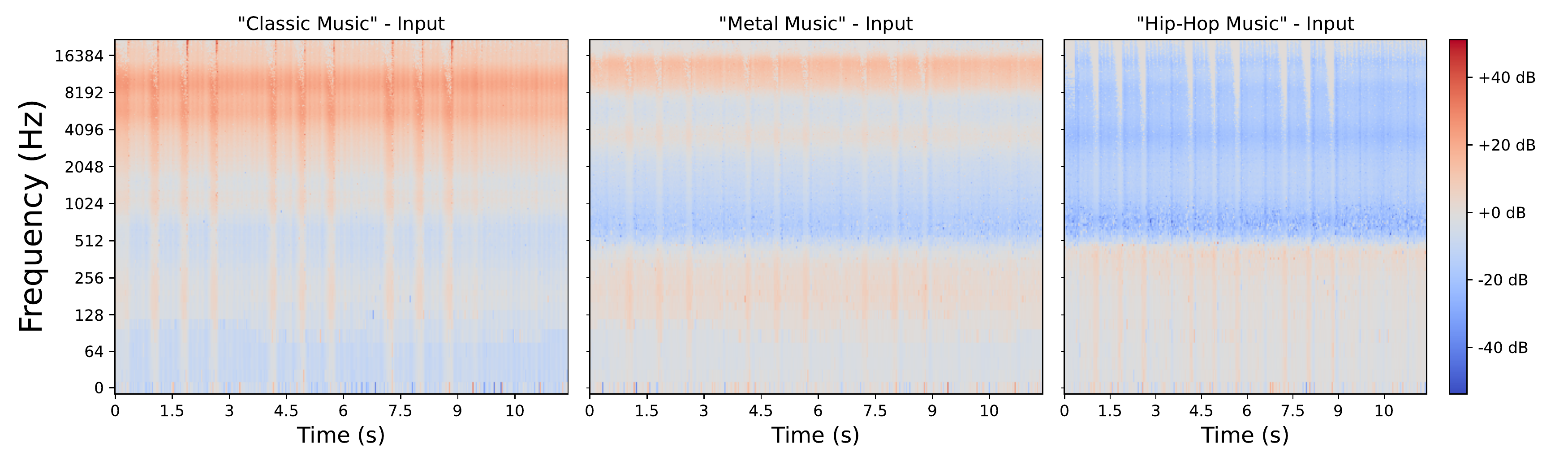}
    }
    \hfill
    \vspace{-0.2cm}
    \subfigure[Spectral Centroid]{
        \includegraphics[width=0.31\textwidth]{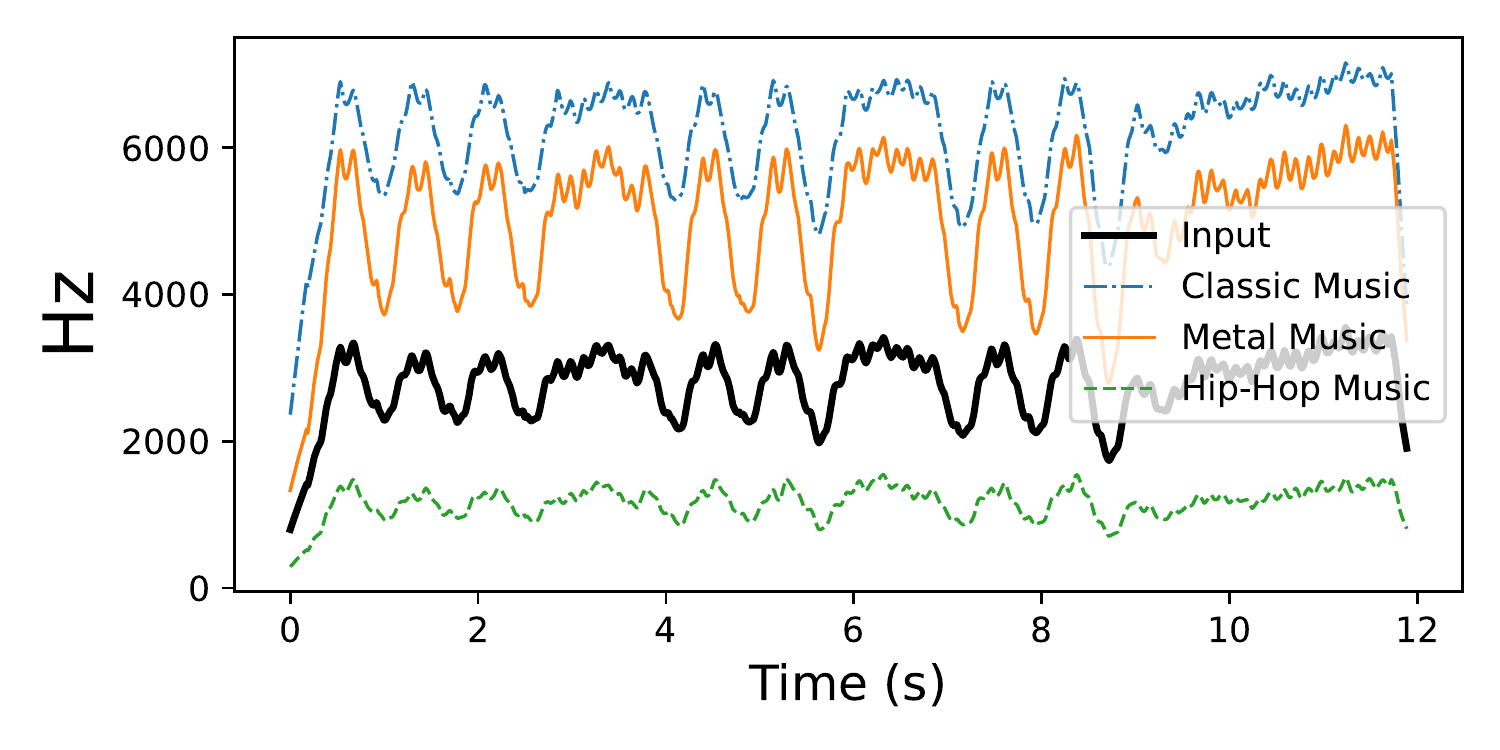}
    }
    \hfill
    \subfigure[Crest Factor]{
        \includegraphics[width=0.31\textwidth]{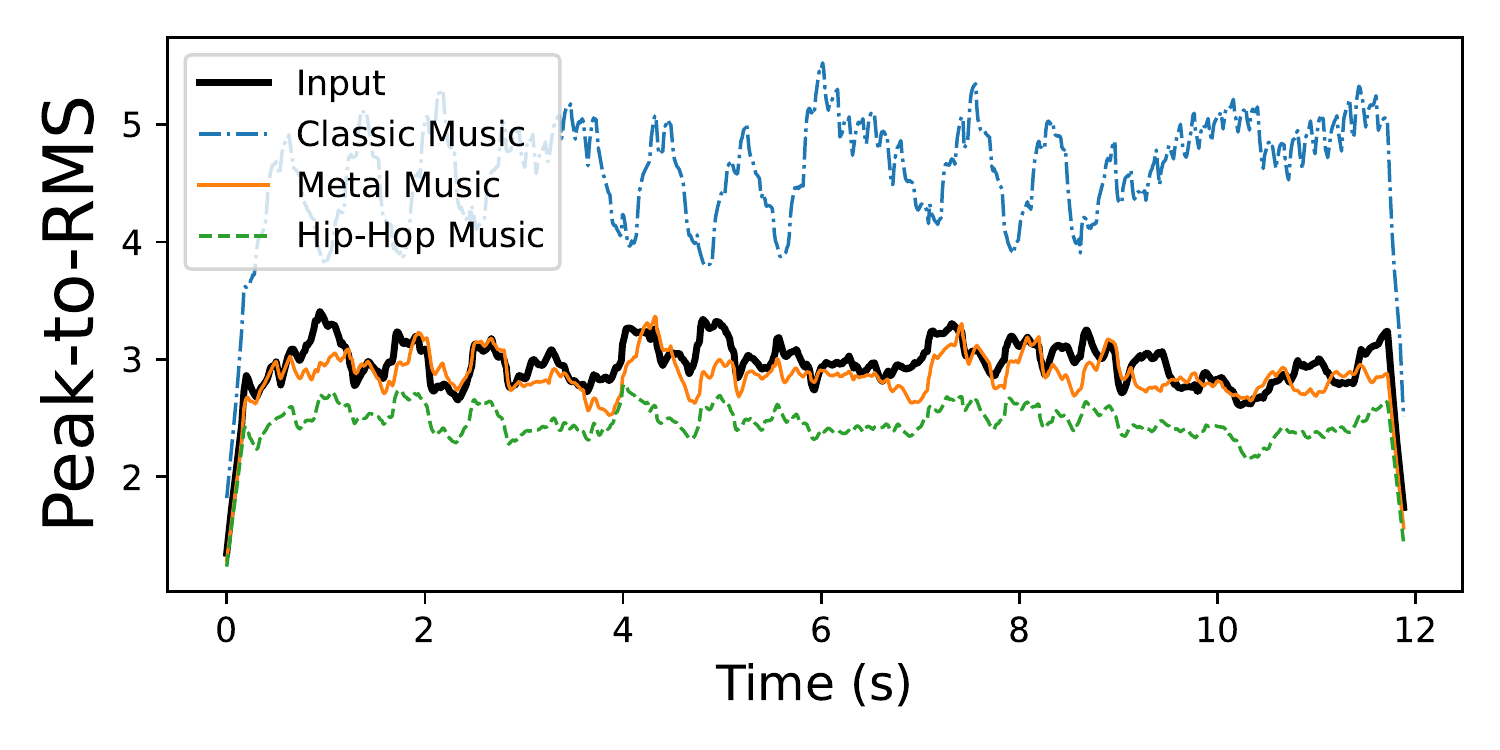}
    }
    \hfill
    \subfigure[RMS Energy]{
        \includegraphics[width=0.31\textwidth]{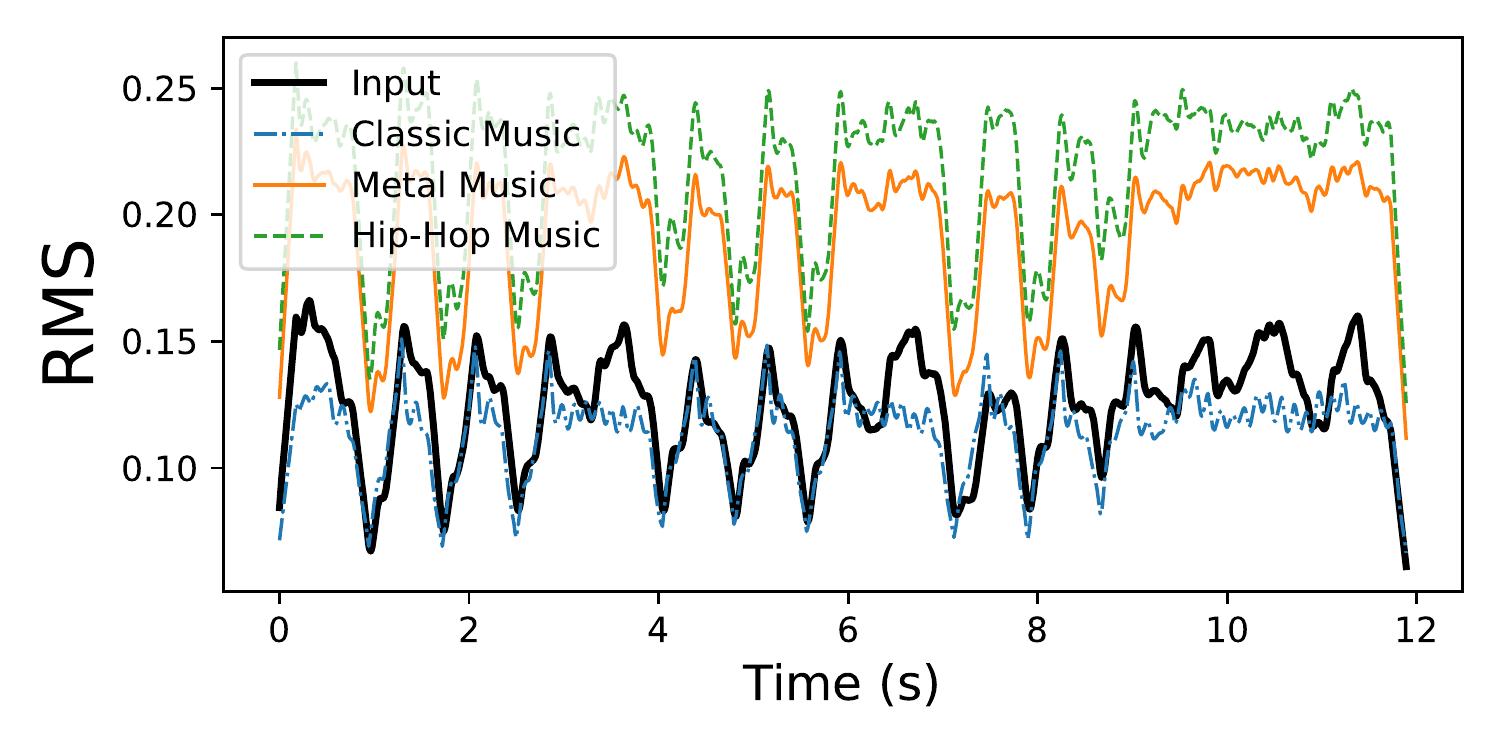}
    }
    
    \vspace{-0.2cm}
    \caption{Comparison of different audio features between the input music and ITO-processed tracks using text prompts \textit{``Classic Music''}, \textit{``Metal Music''}, and \textit{``Hip-Hop Music''}.
    }
    \vspace{-0.2cm}
    \label{fig:qual_analysis}
\end{figure*}

\subsection{Subjective Evaluation}
\label{subsec:sbj_eval}

To further validate our proposed methods subjectively, we conducted a MUSHRA-type listening test with 10 participants, all familiar with music post-production and digital effects, having 2 to 5 years of experience in recording, mixing, or mastering.
Participants rated various processed tracks based on their similarity in mastering audio effects to a reference track. The evaluation included 8 questions, with 30 seconds long music recordings for all stimuli. The reference audio contained different content from the stimuli, but we ensured the reference and stimuli were not too dissimilar in terms of genre or instrumentation. As a low anchor, the initial music track before being inputted into style transfer systems was presented. There was no high anchor, as the evaluation setup aimed to mimic real-world Fx style transfer using music tracks from the Jamendo dataset.

As illustrated in Figure~\ref{fig:subjective_result}, the subjective test results align with the trends observed in our objective evaluations. All our proposed methods surpass the baselines, with results further enhanced by ITO, showing audio effects characteristics more similar to those of the reference. The similarity scores for the proposed system ranged from 0 to 100, indicating that the listening test was highly challenging, even for experts with domain knowledge. Nevertheless, the proposed systems consistently outperformed the baseline, showing significant improvements (pairwise t-test, $p < 0.05$).

\subsection{Qualitative Analysis of ITO with Text Prompts}
\label{subsec:qual_analysis}
To evaluate the effectiveness of ITO under different $\mathcal{L}_\text{aux}$, we perform a qualitative analysis using text prompts with CLAP embeddings~\cite{Laion-CLAP}, similar to the application demonstrated in~\cite{chu2025text2fx}. Given an input music track, we optimize $z_\text{ref}$ of the proposed white-box model using text-based conditioning, leveraging the CLAP embedding cosine similarity as the optimization objective. Specifically, we compute the audio embedding $\text{CLAP}_\text{aud}$ of the steered output and the text embedding $\text{CLAP}_\text{txt}$ of the given reference text prompt, then maximize their cosine similarity to guide the transformation. The input music piece used for this analysis is an 11.8-second-long instrumental rock track. Since the input content remains unchanged across different steered results, we can directly assess the influence of each text prompt on various musical features. We explore ITO with three different prompts: \textit{``Classic Music''}, \textit{``Metal Music''}, and \textit{``Hip-Hop Music''} for analysis.
This experimental setup can be explored through our interactive demo
\footnote{{\fontsize{7.2pt}{7pt}\selectfont\url{https://huggingface.co/spaces/jhtonyKoo/ITO-Master}}}.

As shown in Figure~\ref{fig:qual_analysis}, the optimized results exhibit distinct characteristics that align with general expectations for each genre.
The spectrogram difference plots in~\ref{subfig:qual_spec_diff} (steered output - input) highlight the frequency ranges most affected by ITO. Specifically, the \textit{``Classic Music''} prompt exhibits notable changes in the mid and high frequencies, aligning with the characteristic brightness and clarity often associated with classical recordings. The \textit{``Metal Music''} prompt shows differences in both low and high frequencies, reflecting the genre's typical emphasis on powerful bass and sharp treble for aggressive instrumentation. In contrast, the \textit{``Hip-Hop Music''} prompt predominantly affects the low-frequency range, reinforcing the genre's signature emphasis on deep bass and sub-bass elements, which are essential for driving rhythm-heavy beats.

These observations are further supported by the spectral centroid, crest factor, and RMS energy analyses. The spectral centroid results follow an expected trend, where \textit{hip-hop} has the lowest centroid due to its bass-heavy nature, followed by \textit{metal}, while \textit{classic music} has the highest centroid, reflecting its emphasis on harmonic richness and treble clarity. The crest factor, representing peak-to-RMS ratio, is lowest for \textit{hip-hop}, indicating a more compressed and bass-heavy dynamic structure, while \textit{classic music} has the highest crest factor, aligning with its typically uncompressed, wider dynamic range. RMS energy shows an increasing trend from \textit{classic} to \textit{hip-hop}, with \textit{metal} falling in between, which is consistent with the respective loudness and dynamic characteristics of these genres. These findings suggest that ITO, guided by CLAP$_\text{txt}$, successfully steers the mastering Fx chain to align with the expected sonic characteristics of the given text prompt, demonstrating its potential as a creative tool for music post-production.

\section{Conclusion}
\label{sec:conclusion}



In this paper, we introduced the ITO-Master framework, which leverages ITO on \( z_{\text{ref}} \) for mastering style transfer. Our experiments showed that training the reference encoder \( \Phi \) alongside \( \Psi \) improves performance. Optimizing \( z_{\text{ref}} \) with ITO led to meaningful improvements with few steps, outperforming direct optimization of \( \Theta \) in efficiency. Subjective evaluations confirmed that our method produces perceptually aligned mastering effects, and qualitative results highlighted the potential of text-conditioned ITO for creative applications.
As future work, we plan to incorporate production quality and usability into the evaluation, alongside reference alignment. Since mastering is a curatorial task, poor reference choices can lead to suboptimal results despite high alignment, highlighting the need for perceptual preference metrics.

\bibliography{ISMIRtemplate}

%
%
%
%

\end{document}